\documentclass[aps,pre,twocolumn,showpacs,superscriptaddress,nofootinbib]{revtex4-2}

\usepackage{CJK}
\usepackage{amsfonts,mathrsfs,amsmath,amsthm,amssymb}
\usepackage{newtxmath}
\usepackage{bm,times}
\usepackage{graphicx,epsfig}
\usepackage{tikz,scalerel}
\usepackage[colorlinks=true,breaklinks=true,linkcolor=blue,citecolor=blue,urlcolor=blue]{hyperref}

\hypersetup{
  pdfauthor={Tao Zhou},
  pdfkeywords={},
  pdftitle={Thermodynamic work and heat for a quantum process: Approach by Hamiltonian decomposition},
  pdfsubject={Research Paper},
  pdfpagemode=UseNone
}

\def \tr {\mathrm{Tr}}
\def \non {\nonumber}
\def \d {\mathrm{d}}

\def \bra {\langle}
\def \ket {\rangle}

\newcommand{\be}{\begin{equation}}
\newcommand{\ee}{\end{equation}}
\newcommand{\bes}{\begin{eqnarray}}
\newcommand{\ees}{\end{eqnarray}}

\newcommand{\im}{\mathrm{i}}
\newcommand{\dbar} {\ensuremath{\,\mathchar'26\mkern-12mu \mathrm{d}}}

\renewcommand{\section}[1]{{\par\it #1.---}\ignorespaces}

\newtheorem{lemma}{Lemma}

\makeatletter
    
    \newcommand{\Rmnum}[1]{\expandafter\@slowromancap\romannumeral #1@}
\makeatother

\usetikzlibrary{svg.path}
\definecolor{orcidlogocol}{HTML}{A6CE39}
\tikzset{
	orcidlogo/.pic={
		\fill[orcidlogocol] svg{M256,128c0,70.7-57.3,128-128,128C57.3,256,0,198.7,0,128C0,57.3,57.3,0,128,0C198.7,0,256,57.3,256,128z};
		\fill[white] svg{M86.3,186.2H70.9V79.1h15.4v48.4V186.2z}
		svg{M108.9,79.1h41.6c39.6,0,57,28.3,57,53.6c0,27.5-21.5,53.6-56.8,53.6h-41.8V79.1zM124.3,172.4h24.5c34.9,0,42.9-26.5,42.9-39.7c0-21.5-13.7-39.7-43.7-39.7h-23.7V172.4z}
		svg{M88.7,56.8c0,5.5-4.5,10.1-10.1,10.1c-5.6,0-10.1-4.6-10.1-10.1c0-5.6,4.5-10.1,10.1-10.1C84.2,46.7,88.7,51.3,88.7,56.8z};}}
\newcommand\orcid[1]{\href{https://orcid.org/#1}{\mbox{\scalerel*{\begin{tikzpicture}[yscale=-1,transform shape]\pic{orcidlogo};\end{tikzpicture}}{|}}}}

\begin{document}

\title{Thermodynamic Work and Heat for a Quantum Process: Approach by Hamiltonian Decomposition}
\author{Tao Zhou\orcid{0000-0001-7069-9505}}
\email{taozhou@swjtu.edu.cn}
\affiliation{Quantum Optoelectronics Laboratory, School of Physical Science and Technology, Southwest Jiaotong University, Chengdu 610031, China}
\affiliation{Department of Applied Physics, School of Physical Science and Technology, Southwest Jiaotong University, Chengdu 611756, China}

\author{Jiangyang Pu}
\affiliation{Quantum Optoelectronics Laboratory, School of Physical Science and Technology, Southwest Jiaotong University, Chengdu 610031, China}

\author{Xiaohua Wu}
\affiliation{College of Physical Science and Technology, Sichuan University, Chengdu 610064, China}

\date{\today}

\begin{abstract}
The separation of internal energy into heat and work in quantum thermodynamics is a controversial issue for a long time, and we revisit and solve this problem in this work. It is shown that the Hamiltonian plays dual roles for a quantum system, and by decomposing the interaction Hamiltonian between system and environment accordingly, an ``effective Hamiltonian" for an open quantum system can be proposed. The explicit expression of the effective Hamiltonian is obtained systematically, and as a consequence, the internal energy of an open quantum system can be well defined, leading to the reasonable definitions of  work and heat for a general quantum process.
\end{abstract}


\maketitle

\section{Introduction}
\label{Sec1}
Thermodynamics in quantum regime is an emerging field lying at the intersection of thermodynamics and quantum mechanics. It seeks to understand the fundamental principles of thermodynamics at quantum level, and plays an important role in quantum information science, statistic mechanics, and quantum technologies~\cite{gemmer2004quantum,deffner2019quantum,RevModPhys.81.1665,RevModPhys.83.771,Goold_2016}. Quantum thermodynamics is different from classical thermodynamics in the emphasis on the dynamical process out of equilibrium. The laws of quantum thermodynamics have attracted much attention~\cite{PhysRevLett.93.140403,pnas.1411728112,WOS:000396104200001,PhysRevE.102.062152,e15062100}, and recently, the influence of quantum coherence and entanglement on the laws of thermodynamics has been widely discussed~\cite{PhysRevE.102.062152,PhysRevLett.113.150402,e15062100,PhysRevE.99.042105,WOS:000459897600125,PhysRevX.5.021001,RevModPhys.89.041003,PhysRevA.79.010101,WOS:000462340700001,WOS:000444757900010,Brandner_2015,Korzekwa_2016}.

Quantum thermodynamics has an intimate connection with the theory of open quantum systems~\cite{10.1093/acprof:oso/9780199213900.001.0001}, and there is a long-standing open problem that the consistent definitions of work, heat, temperature, and entropy production for a nonequilibrium evolution of an open quantum system coupled to environment are lacked. Although extensive efforts have been put on this topic, some of the results are not consistent with each other, and lead to unexpected conclusions according to thermodynamics. This topic is still controversial~\cite{RevModPhys.93.035008,Weimer_2008,Esposito_2010,PhysRevE.83.041131,Alipour2016,PhysRevX.7.021003,PhysRevA.98.012130,PhysRevE.99.012120,PhysRevLett.124.160601,PhysRevA.105.L040201,PhysRevA.105.052216}, and for instance, there are semiclassical approaches to the definitions of the work and heat~\cite{Bender_2000,PhysRevE.83.041117,POLKOVNIKOV2011486,Kieu2006,PhysRevLett.93.140403,PhysRevE.76.031105}, while the dynamical master equation with coherent and dissipative parts is widely employed, where the internal energy changes due to the coherent (dissipative) part is defined as work (heat)~\cite{Alicki_1979,10.1063/1.523789}. 

The key obstacle to a satisfactory definition of work and heat in quantum thermodynamics lies in two aspects: One is that the thermodynamic variables are not observables described by Hermitian operators~\cite{PhysRevE.75.050102} and trajectory-dependent quantities~\cite{PhysRevLett.100.020601,C6SC04350J,Niedenzu2019conceptsofworkin,PhysRevA.97.012131,Elouard2017}, and the other is that the definition of internal energy of an open quantum system is still absent up to now~\cite{PhysRevA.108.042209}. To clarify this issue, the energy exchange between an open quantum system $S$ and the environment $E$ through the interaction should be taken into account. If the quantum system $SE$ consisting of the open system and the environment is closed, both the system $S$ and the environment $E$ may be described by the quantum master equation.  According to the double roles of Hamiltonian, a true effective Hamiltonian can be obtained by separating the Hamiltonian of the whole system $SE$, and the work and heat in quantum thermodynamics can be well defined.

\section{Dual roles of the Hamiltonian and the closed quantum system revisited}
\label{Sec2}
For a closed quantum system driven by a time-dependent Hamiltonian $H(t)$, its internal energy of the system can be defined as
\be
\label{InE}
U(t)=\tr\left[\rho(t)H(t)\right].
\ee
with $\rho(t)$ the density operator. Then, one can have the variation of the internal energy
\be
\label{Change}
\d U=\tr\left(\d\rho H\right)+\tr\left(\rho\d H\right)\equiv\dbar Q+\dbar W,
\ee
where $\dbar Q\equiv\tr\left(\d\rho H\right)$ and $\dbar W\equiv\tr\left(\rho\d H\right)$ are defined as the heat change and work change, respectively~\cite{PhysRevLett.93.140403,Alicki_1979}, since the changes $\tr\left(\d\rho H\right)$ and $\tr\left(\rho\d H\right)$ derive from the changes of the state $\rho(t)$ and the Hamiltonian $H(t)$, respectively. Meanwhile, the evolution of the system is described by the Liouville-von Neumann equation
\be
\label{Evo}
\dot{\rho}(t)=-\im\left[H(t),\rho(t)\right]
\ee
where the overdot $\dot{\bullet}$ and $\left[\bullet,\bullet\right]$ represent the time derivative and the commutator, respectively, and we have set $\hbar=1$. With Eq.~\eqref{Evo}, $\dbar Q=-\im\tr\left([H,\rho]H\right)\d t=-\im\tr\left(H\rho H-\rho HH\right)\d t=0$, showing that there is no heat exchange for a unitary evolution. It is obvious from Eq.~\eqref{InE} and Eq.~\eqref{Evo} that the Hamiltonian $H(t)$ plays double roles for the closed quantum system: One is that the internal energy is the expectation of $H(t)$ with respect to $\rho(t)$, and the other is that the evolution is generated by $H(t)$. 

To have an in-depth insight into the Hamiltonian, the following lemma is useful.
\begin{lemma}
\label{Lem1}
For a closed quantum system described by the density operator $\rho(t)$, its Hamiltonian $H(t)$ can be decomposed as $H(t)=\bar{H}(t)+H^\bot(t)$ where $\tr\left[\bar{H}(t)H^\bot(t)\right]=0$, such that $\left[\bar{H}(t),\rho(t)\right]=0$ and $\tr\left[H^\bot(t)\rho(t)\right]=0$.
\end{lemma}
{\it Proof.} First, the density operator $\rho(t)$ has a spectral decomposition $\rho(t)=\sum_{i=1}^d\lambda_i(t)\left|\phi_i(t)\right\ket\left\bra\phi_i(t)\right|$, where $d$ is the dimension of the Hilbert space $\mathcal{H}$ of the system. One can introduce the transition-projection operators $P_{ij}=|\phi_i\ket\bra\phi_j|$ to construct the $d^2$ generators of Lie algebra $\mathfrak{u}(d)=\left\{A\in\mathrm{GL}(d,\mathbb{C})|A^\dag=A\right\}$ of the $\mathrm{U}(d)$ group as follows
\bes
&&\openone=\sum_{i=1}^d P_{ii},\ \ w_l=-\sqrt{\frac{2}{l(l+1)}}\left(\sum_i^lP_{ii}-lP_{l+1,l+1}\right),\label{Proj}\\
&&u_{ij}=P_{ij}+P_{ji},\ \ v_{ij}=\im\left(P_{ij}-P_{ji}\right),\label{Tran}
\ees
where $1\leqslant l\leqslant d-1$ and $1\leqslant i<j\leqslant d$. Except the identity operator $\openone$, the other generators above are all traceless, and can be selected as $d^2-1$ generators of Lie algebra $\mathfrak{su}(d)=\left\{A\in\mathrm{GL}(d,\mathbb{C})|A^\dag=A,\tr A=0\right\}$~\cite{PhysRevA.52.4396}.

Second, one can introduce the Hilbert-Schmidt scalar product $(A,B)=\tr\left(A^\dag B\right)$ on the Lie algebra $\mathfrak{u}(d)$ and $\mathfrak{su}(d)$, and the generators above are all orthogonal~\cite{PhysRevA.52.4396}. The generators in Eq.~\eqref{Proj} constructed by the projection operators all commute with $\rho(t)$, and can span a subspace $\bar{\mathfrak{h}}$ of $\mathfrak{u}(d)$. Meanwhile, the generators in Eq.~\eqref{Tran} constructed by the transition operators are all orthogonal to $\rho(t)$, and can span another subspace $\mathfrak{h}^\bot$. Obviously, $\mathfrak{u}(d)$ has a decomposition
\be
\label{DecAlg}
\mathfrak{u}(d)=\bar{\mathfrak{h}}\oplus\mathfrak{h}^\bot.
\ee

Finally, both $\rho(t)$ and $H(t)$ are Hermitian operators on the Hilbert space $\mathcal{H}$, and $\rho(t)\in \bar{\mathfrak{h}}$ and $H(t)\in \mathfrak{u}(d)$, respectively. Denote the projection on $\bar{\mathfrak{h}}$ by $\Pi$, and one can obtain
\bes
\label{HamProj}
\bar{H}(t)&=&\Pi\left[H(t)\right],\ \ H^\bot(t)=H(t)-\Pi\left[H(t)\right],\\
H(t)&=&\bar{H}(t)+H^\bot(t),\label{HamProj}
\ees
which satisfy $\left[\bar{H}(t),\rho(t)\right]=0$ and $\tr\left[H^\bot(t)\rho(t)\right]=0$, respectively. \qed

According to the Lemma above, the internal energy can be further expressed as
\be
\label{InE2}
U(t)=\tr\left[\rho(t)\bar{H}(t)\right],
\ee
and the evolution of the system can be described by
\be
\label{Evo2}
\dot{\rho}(t)=-\im\left[H^\bot(t),\rho(t)\right].
\ee
The Eqs.~\eqref{InE2} and~\eqref{Evo2} above imply that the two Hamiltonians $\bar{H}(t)$ and $H^\bot(t)$ are responsible for the double roles of $H(t)$ respectively: On the one hand, the internal energy is the expectation of $\bar{H}(t)$ with respect to $\rho(t)$, and on the other hand, the evolution of the system is governed by $H^\bot(t)$. Thus, the heat change and work change in Eq.~\eqref{Change} can be expressed as
\be
\dbar Q=\tr\left(\d\rho\bar{H}\right),\ \ \dbar W=\tr\left(\rho\d\bar{H}\right).
\ee
With Eq.~\eqref{Evo2}, one can also have $\dbar Q=0$ for the unitary process.

\section{Splitting the interaction Hamiltonian and solution for the open quantum system}
\label{Sec3}
An open quantum system interacts with its environment, leading to extremely different dynamical properties compared with a closed one. For technical simplicity, we consider a total system $SE$ consisting of two subsystems with interaction: the open quantum system of interest $S$ and the environment $E$, and the density operator for the total system is $\rho_{SE}(t)$. It can always be assumed that the total system has no interaction with other systems, and then the system $SE$ is closed and governed by a autonomous Hamiltonian
\be
\label{THam}
H=H_S\otimes \openone_E+\openone_S\otimes H_E+H_I,
\ee
where the free Hamiltonians $H_S$ and $H_E$ of the system $S$ and the environment $E$, respectively, and the interaction Hamiltonian $H_I$ are all time-independent, and $\openone_S$ ($\openone_E$) is the identity operator for the system $S$ (environment $E$).

According to the Hamiltonian in Eq.~\eqref{THam}, the internal energy of $SE$ is
\be
\label{TInE}
U_{SE}=\tr\left(H_S\rho_S\right)+\tr\left(H_E\rho_E\right)+\tr\left[H_I\rho_{ES}(t)\right],
\ee
where $\rho_S=\tr_E\rho_{SE}$ and $\rho_E=\tr_S\rho_{SE}$ denote the reduced density operators of the system and environment, respectively. The express in Eq.~\eqref{TInE} has caused great confusion in the definition of internal energy $U_S$ of the open quantum system $S$. First, the internal energy of subsystem $S$ is defined by the expectation of its ``free Hamiltonian" $U_S=\tr\left(H_S\rho_S\right)$ in some literatures~\cite{PhysRevLett.116.240403,PhysRevE.104.044111}. Second, the term $U_I=\tr\left[H_I\rho_{SE}(t)\right]$ contributing to the total internal energy is introduced as “binding energy”, resulting in $U_S+U_E<U_{SE}=\tr\left[H\rho_{SE}(t)\right]$.  Consequently, the definitions of work and heat in quantum thermodynamics become much more complicated.

The free Hamiltonians $H_S$ and $H_E$ in Eq.~\eqref{THam} drive the local systems $S$ and $E$ unitarily. The interaction $H_I$ drive a unitary evolution on the total system, and hence usually generates both unitary and dissipative evolutions on both subsystems, which leading to the work and the heat exchanges between $S$ and $E$, respectively. Illuminated by the fact that the dual roles of Hamiltonian can be undertaken by two splitted ones for a unitary evolution, we can observe something about it in hindsight. The interaction Hamiltonian $H_I$ can be decomposed as $H_I=H'_S(t)\otimes\openone_E+\openone_S\otimes H'_E(t)+H'_I(t)$, conditioned on the following requirements: (i) $H'_I$ generates only dissipations and thus no unitary evolution on subsystems, and $H'_S(t)$ and $H'_E(t)$ govern the unitary evolutions generated by the interaction $H_I$; (ii) The ``effective Hamiltonians" $H_S^{\mathrm{Eff}}(t)=H_S+H'_S(t)$ and $H_E^{\mathrm{Eff}}(t)=H_E+H'_E(t)$ for subsystems including the interaction effects can be introduced, and the internal energy is the average of the two corrected observables $U_S=\tr\left(H_S^\mathrm{Eff}\rho_S\right)$ and $U_E=\tr\left(H_E^\mathrm{Eff}\rho_E\right)$~\cite{Weimer_2008,Alipour2016,PhysRevLett.124.160601,PhysRevA.105.052216}; (iii) It is further required that $\tr\left(H'_I\rho_{SE}\right)=0$ such that the binding energy is zero and  the total internal energy is the sum of subsystems' $U_S(t)+U_E(t)=U_{SE}$.

To derive the explicit expressions for $H'_S(t)$, $H'_E(t)$, and $H'_I(t)$, one can consider the dynamics of the system $S$
\bes
\label{ReDyn}
\dot{\rho}_S&=&-\im\tr_E\left[H,\rho_{SE}\right]\non\\
&=&-\im\left[H_S,\rho_S\right]-\im\tr_E\left[H_I,\rho_{SE}\right],
\ees
where $-\im\left[H_S,\rho_S\right]$ represents the unitary evolution generated by $H_S$, and $\mathbb{H}_S\equiv-\im\tr_E\left[H_I,\rho_{SE}\right]=\sum_{ij}\left(\mathbb{H}_S\right)_{ij}|\phi_i^S\ket\bra\phi^S_j|$ represents the dynamics generated by $H_I$ including both the unitary and dissipative ones, with $|\phi_i^S\ket$ the eigenvectors of $\rho_S$, say $\rho_S(t)=\sum_i\lambda^S_i(t)|\phi_i^S(t)\ket\bra\phi_i^S(t)|$, $\lambda^S_i>0$, and $\left(\mathbb{H}_S\right)_{ij}=\bra\phi^S_i|\mathbb{H}_S|\phi_j^S\ket$. One can also construct a Lie algebra $\mathfrak{u}(d_S)=\bar{\mathfrak{h}}_S\oplus\mathfrak{h}_S^\bot$ as in Eq.~\eqref{Proj} and Eq.~\eqref{Tran}, where $d_S$ is the rank of $\rho_S(t)$. Now, $H'_S$ can be further decomposed as $H'_S=\bar{H}'_S+{H'}_S^\bot$, with $\bar{H}'_S\in \bar{\mathfrak{h}}_S$ and ${H'}^\bot_S\in \mathfrak{h}_S^\bot$. Due to the facts $\left[|\phi_i^S\ket\bra\phi_j^S|,\rho_S\right]=\lambda_{ji}|\phi_i^S\ket\bra\phi_j^S|$ and $\left[|\phi_i^S\ket\bra\phi_i^S|,\rho_S\right]=0$, with $\lambda^S_{ji}=\lambda^S_j-\lambda^S_i$ the difference between the eigenvalues of $|\phi_j^S\ket$ and $|\phi_i^S\ket$, one can obtain 
\be
\label{H'bot}
{H'}_S^\bot=\im\sum_{i\neq j}\frac{\left(\mathbb{H}_S\right)_{ij}}{\lambda_{ij}}|\phi_i^S\ket\bra\phi_j^S|.
\ee
Similarly, ${H'}_E^\bot$ can also be obtained. Up to now, all the eigenvalues $\lambda^S_i(t)$ are assumed to be non-degenerate, and the extension to the degenerate case is addressed in the subsequent section.

The constraint $\tr\left(H'_I\rho_{SE}\right)=0$ only requires that
\be
\tr\left(\rho_{SE}H'\right)=\tr\left(\rho_{SE}H_I\right)=U_I,
\ee
where $H'=H'_S\otimes\openone_E+\openone_S\otimes H'_E$, and therefore, to derive $\bar{H}'_S$ and $\bar{H}'_E$, more constraints are necessary. It is reasonable to further require that the distance between the two interaction Hamiltonians $H_I$ and $H'_I$
\be
d=\left\|H_I-H'_I\right\|=\left\|H'\right\|\equiv \sqrt{\tr H^{\prime2}}
\ee
attains the minimum value, and with the Lagrangian multiplier method, one can come to
\be
\bar{H}'_S=\sum_{i=1}^{r_S}\frac{D_1U_I}{D_2}|\phi_i^S\ket\bra\phi_i^S|,
\ee
where
\bes
D_1&=&\frac{1}{2}\left(\frac{d_E-1}{d_Sd_E-1}-\lambda^S_i\right),\non\\
D_2&=&\frac{d_S+d_E-2}{2(d_Sd_E-1)}-\sum_{i=1}^{r_S}(\lambda_i^S)^2-\sum_{j=1}^{r_E}(\lambda_j^E)^2,\non
\ees
with $d_S(d_E)$ and $\lambda_i^S(\lambda_j^E)$ the rank and eigenvalues of system (Environment). A similar expression can be obtained for $\bar{H}'_E$, but not displayed here.

Now, one can obtain the expression for system's effective Hamiltonian $H_S^{\mathrm{Eff}}=H_S+H'_S$, and the internal energy for system is
\be
U_S(t)=\tr\left[\rho_S(t)H_S^{\mathrm{Eff}}(t)\right].
\ee
Heat exchange is
\be
\dbar Q=\tr\left[\d\rho_S(t)H_S^{\mathrm{Eff}}(t)\right]=\sum_i\left(\mathbb{H}_S\right)_{ii}\bra\phi^S_i|H_S^{\mathrm{eff}}|\phi_i^S\ket,
\ee
and work exchange is
\be
\dbar W=\tr\left[\rho_S(t)\d H_S^{\mathrm{Eff}}(t)\right]=\tr\left[\rho_S(t)\d H'_S(t)\right].
\ee

\section{Remarks and discussions}
\label{Sec4}
(i) A statement is worthy mentioning to avoid confusion. In this work, a system is called (a) isolated, (b) closed, and (c) open, if (a) it has no interaction and exchanges no work and heat with its surroundings,  (b) it evolves unitarily, i.e., it may exchange work with its surroundings driven by some classical field, and (c) it interacts with its surroundings and evolves non-unitarily, exchanging both work and heat with the surroundings.

(ii) The generators in Eq.~\eqref{Proj} and Eq.~\eqref{Tran} are constructed with the transition-projection operators $P_{ij}=|\phi_i\ket\bra\phi_j|$. Since for any arbitrary real function $\theta_i(t)$, $\left|\phi'_i(t)\right\ket=e^{\im\theta_i(t)}\left|\phi_i(t)\right\ket$ can also be the eigenvectors of $\rho(t)$, one may wonder whether the decomposition in Eq.~\eqref{DecAlg} is unique. To clarify this, the new projection-transition operators $P'_{ij}=|\phi'_i\ket\bra\phi'_j|$ can be introduced. Since $P'_{ii}=|\phi'_i\ket\bra\phi'_i|=|\phi_i\ket\bra\phi_i|=P_{ii}$, the subspace $\bar{\mathfrak{h}}$ constructed by the projection operators are the same. For the generators constructed by the transition operators, one can have
\bes
u'_{ij}&=&P'_{ij}+P'_{ji}=u_{ij}\cos\theta_{ij}-v_{ij}\sin\theta_{ij},\non\\
v'_{ij}&=&\im\left(P'_{ij}-P'_{ji}\right)=-u_{ij}\sin\theta_{ij}+v_{ij}\cos\theta_{ij},\non
\ees
with $\theta_{ij}=\theta_i-\theta_j$, which shows that the subspace $\mathfrak{h}^\bot$ constructed by the transition operators $u'_{ij}$ and $v'_{ij}$ are the same as that constructed by $u_{ij}$ and $v_{ij}$. Therefore, under the transformation
\be
\label{GauTrans}
\left|\phi_i(t)\right\ket\rightarrow\left|\phi'_i(t)\right\ket=e^{\im\theta_i(t)}\left|\phi_i(t)\right\ket,
\ee
the decomposition in Eq.~\eqref{DecAlg} is unique.

(iii) According to Lemma~\ref{Lem1}, the evolution of state $\rho(t)$ for a closed system is governed by $H^\bot(t)$ only as in Eq.~\eqref{Evo2}. However, the eigenvectors $\left|\phi_i(t)\right\ket$ evolves as
\be
\im|\dot{\phi}_i(t)\ket=H|\phi_i(t)\ket,\non
\ee
showing that the evolution of $\left|\phi_i(t)\right\ket$ is governed by both $\bar{H}(t)$ and $H^\bot(t)$. If $\bar{H}(t)=0$, one can have $\bra\phi_i(t)|\dot{\phi}_i(t)\ket=-\im\bra\phi_i(t)|H^\bot(t)|\phi_i(t)\ket=0$, which are the parallel transport conditions proposed in Refs.~\cite{PhysRevA.67.032106,PhysRevLett.93.080405}. Under these conditions, the dynamic phase is zero, rendering the geometric phase the total phase in the evolution. Thus, it is implied that the Hamiltonians $\bar{H}(t)$ and $H^\bot(t)$ have deep connections with dynamical and geometric phases, respectively. To figure out the relationship is a nontrivial task, and will be our next work.

(iv) When the state $\rho_S(t)$ of the system is degenerate, the difference $\lambda^S_{ij}$ in Eq.~\eqref{H'bot} may be zeros and thus the Hamiltonian ${H'}_S^\bot$ is {\it ad hoc} and required to be modified. Consider the case
\be
\rho_S(t)=\sum_k\sum_{\mu=1}^{n_k}\lambda^S_k(t)\left|\phi_S^{k\mu}(t)\right\ket\left\bra\phi_S^{k\mu}(t)\right|,\text{with}\ t\in[t_0,t_1],\non
\ee
where $\lambda^S_k(t)$ are the eigenvalues of $\rho_S(t)$ with degeneracy $n_k$, and $|\phi_S^{k\mu}(t)\ket,\ \mu=1,2,...,n_k$ are the corresponding eigenvectors. The $k$th eigenspace $\mathcal{H}_S^k$ is a subspace spanned by $\left\{|\phi_S^{k\mu}(t)\ket\big|\mu=1,2,...,n_k\right\}$, and then the system's Hilbert space $\mathcal{H}_S=\bigoplus_k\mathcal{H}_S^k$. To derive a proper ${H'}_S^\bot$, appropriate eigenvectors $|\tilde{\phi}_S^{k\mu}(t)\ket$ can be chosen such that $\rho_S(t)=\sum_k\sum_{\mu=1}^{n_k}\lambda^S_k(t)|\tilde{\phi}_S^{k\mu}(t)\ket\bra\tilde{\phi}_S^{k\mu}(t)|$ and the operator
\bes
\mathbb{H}_S&=&\sum_k\sum_{\mu=1}^{n_k}\left(\mathbb{H}_S\right)^{kk}_{\mu\mu}\left|\tilde{\phi}_S^{k\mu}(t)\right\ket\left\bra\tilde{\phi}_S^{k\mu}(t)\right|\non\\
&&+\sum_{k\neq l}\sum_{\mu=1}^{n_k}\sum_{\nu=1}^{n_l}\left(\mathbb{H}_S\right)^{kl}_{\mu\nu}\left|\tilde{\phi}_S^{k\mu}(t)\bigg\ket\bigg\bra\tilde{\phi}_S^{l\nu}(t)\right|\non
\ees
is diagonal on each subspace $\mathcal{H}_S^k$, where $\left(\mathbb{H}_S\right)^{kl}_{\mu\nu}=\bra\tilde{\phi}_S^{k\mu}|\mathbb{H}_S|\tilde{\phi}_S^{l\nu}\ket$. Therefore, one can obtain
\be
{H'}_S^\bot=\im\sum_{k\neq l}\sum_{\mu=1}^{n_k}\sum_{\nu=1}^{n_l}\frac{\left(\mathbb{H}_S\right)^{kl}_{\mu\nu}}{\lambda^S_{kl}}\left|\tilde{\phi}_S^{k\mu}(t)\bigg\ket\bigg\bra\tilde{\phi}_S^{l\nu}(t)\right|.\non
\ee

(v) When the quantum process of the subsystem $S$ can be described by the time-local master equation $\dot{\rho}_S=-\im[K_S(t),\rho_S(t)]+\mathcal{D}[\rho_S(t)]$, the Hermitian operator $K_S(t)$ is conceived as the effective Hamiltonian, and employed to construct the exact thermodynamical quantities, which gives a possible way to define the internal energy, work and heat. However, this prescription is problematic as the splitting of the master equation into unitary part $-\im[K_S(t),\rho_S(t)]$ and dissipative part $\mathcal{D}[\rho_S(t)]$ is highly nonunique~\cite{PhysRevA.105.052216,PhysRevA.108.022209}. Moreover, even for the dissipative part $\mathcal{D}[\rho_S(t)]$, it may still generate unitary dynamics for certain density operators $\rho_S(t)$~\cite{Tobe}.

(vi) The density operator of the total system $SE$ can be expressed as $\rho_{SE}=\rho_S\otimes\rho_E+C$, where $C$ is the operator describing the correlation between system and environment. Then, in Eq.~\eqref{ReDyn}, the term $\mathbb{H}_S=-\im\tr_E\left[H_I,\rho_{SE}\right]$ representing the dynamics of system $S$ generated by $H_I$ can be further expressed as~\cite{Weimer_2008}
\be
\mathbb{H}_S=-\im\left[H''_S,\rho_S\right]-\im\tr_E\left[H_I,C\right],\non
\ee
where $H''_S=\tr_E\left[H_I\left(\openone_S\otimes\rho_E\right)\right]$ is a Hermitian operator, revealing that the interaction $H_I$ generates no dissipative dynamics when $\rho_{SE}$ is a product state ($C=0$). For the second term $-\im\tr_E\left[H_I,C\right]$, it is also Hermitian and usually cannot be diagonal in the basis consisting of eigenvectors of $\rho_S$, which means that this term may generate unitary dynamics. From our discussion, it is not the truth that the process generated by $-\im\tr_E\left[H_I,C\right]$ cannot result in unitary dynamics.

(vii) In this work, the effective Hamiltonian $H_S^{\mathrm{Eff}}$ is dependent on the free Hamiltonian $H_S$, the interaction Hamiltonian $H_I$, and the state $\rho_{SE}$ of the total system. Therefore, the internal energy $U_S$ of system $S$ is also dependent on $H_S$, $H_I$, and $\rho_{SE}$. This result coincides with the conclusion in Ref.~\cite{PhysRevA.108.042209} that the internal energy of an open quantum system $S$ cannot be a functional of only $\rho_S$ and $\dot{\rho}_S$.

\section{Conclusions and summaries}
\label{Sec5}
The assignment of energy of an open quantum system and environment has been revisited in this work. Usually, there are many ways of separating the internal energy of the system from that of the environment, and simultaneously, it is also a difficulty  to distinguish the work from the heat in a quantum process. According to the dual roles of Hamiltonian, the effective Hamiltonian of an open quantum system is proposed by Hamiltonian decomposition. The thermodynamic quantities such as internal energy, work and heat are well defined for quantum system, which are all dynamics-dependent. We expect that the results in this work could lead to further theoretical or experimental consequences.

\section{Acknowledgements}
This work was supported by the National Natural Science Foundation of China (Grant No.~12147208), and the Fundamental Research Funds for the Central Universities (Grant No. 2682021ZTPY050).

\bibliography{refs.bib}

\begin{thebibliography}{53}%
\makeatletter
\providecommand \@ifxundefined [1]{%
 \@ifx{#1\undefined}
}%
\providecommand \@ifnum [1]{%
 \ifnum #1\expandafter \@firstoftwo
 \else \expandafter \@secondoftwo
 \fi
}%
\providecommand \@ifx [1]{%
 \ifx #1\expandafter \@firstoftwo
 \else \expandafter \@secondoftwo
 \fi
}%
\providecommand \natexlab [1]{#1}%
\providecommand \enquote  [1]{``#1''}%
\providecommand \bibnamefont  [1]{#1}%
\providecommand \bibfnamefont [1]{#1}%
\providecommand \citenamefont [1]{#1}%
\providecommand \href@noop [0]{\@secondoftwo}%
\providecommand \href [0]{\begingroup \@sanitize@url \@href}%
\providecommand \@href[1]{\@@startlink{#1}\@@href}%
\providecommand \@@href[1]{\endgroup#1\@@endlink}%
\providecommand \@sanitize@url [0]{\catcode `\\12\catcode `\$12\catcode
  `\&12\catcode `\#12\catcode `\^12\catcode `\_12\catcode `\%12\relax}%
\providecommand \@@startlink[1]{}%
\providecommand \@@endlink[0]{}%
\providecommand \url  [0]{\begingroup\@sanitize@url \@url }%
\providecommand \@url [1]{\endgroup\@href {#1}{\urlprefix }}%
\providecommand \urlprefix  [0]{URL }%
\providecommand \Eprint [0]{\href }%
\providecommand \doibase [0]{https://doi.org/}%
\providecommand \selectlanguage [0]{\@gobble}%
\providecommand \bibinfo  [0]{\@secondoftwo}%
\providecommand \bibfield  [0]{\@secondoftwo}%
\providecommand \translation [1]{[#1]}%
\providecommand \BibitemOpen [0]{}%
\providecommand \bibitemStop [0]{}%
\providecommand \bibitemNoStop [0]{.\EOS\space}%
\providecommand \EOS [0]{\spacefactor3000\relax}%
\providecommand \BibitemShut  [1]{\csname bibitem#1\endcsname}%
\let\auto@bib@innerbib\@empty
\bibitem [{\citenamefont {Gemmer}\ \emph {et~al.}(2004)\citenamefont {Gemmer},
  \citenamefont {Michel},\ and\ \citenamefont {Mahler}}]{gemmer2004quantum}%
  \BibitemOpen
  \bibfield  {author} {\bibinfo {author} {\bibfnamefont {J.}~\bibnamefont
  {Gemmer}}, \bibinfo {author} {\bibfnamefont {M.}~\bibnamefont {Michel}},\
  and\ \bibinfo {author} {\bibfnamefont {G.}~\bibnamefont {Mahler}},\
  }\href@noop {} {\emph {\bibinfo {title} {Quantum Thermodynamics}}}\ (\bibinfo
   {publisher} {Springer, Berlin},\ \bibinfo {year} {2004})\BibitemShut
  {NoStop}%
\bibitem [{\citenamefont {Deffner}\ and\ \citenamefont
  {Campbell}(2019)}]{deffner2019quantum}%
  \BibitemOpen
  \bibfield  {author} {\bibinfo {author} {\bibfnamefont {S.}~\bibnamefont
  {Deffner}}\ and\ \bibinfo {author} {\bibfnamefont {S.}~\bibnamefont
  {Campbell}},\ }\href@noop {} {\emph {\bibinfo {title} {Quantum
  Thermodynamics: An introduction to the thermodynamics of quantum
  information}}}\ (\bibinfo  {publisher} {Morgan \& Claypool Publishers},\
  \bibinfo {year} {2019})\BibitemShut {NoStop}%
\bibitem [{\citenamefont {Esposito}\ \emph {et~al.}(2009)\citenamefont
  {Esposito}, \citenamefont {Harbola},\ and\ \citenamefont
  {Mukamel}}]{RevModPhys.81.1665}%
  \BibitemOpen
  \bibfield  {author} {\bibinfo {author} {\bibfnamefont {M.}~\bibnamefont
  {Esposito}}, \bibinfo {author} {\bibfnamefont {U.}~\bibnamefont {Harbola}},\
  and\ \bibinfo {author} {\bibfnamefont {S.}~\bibnamefont {Mukamel}},\
  }\bibfield  {title} {\bibinfo {title} {Nonequilibrium fluctuations,
  fluctuation theorems, and counting statistics in quantum systems},\ }\href
  {https://doi.org/10.1103/RevModPhys.81.1665} {\bibfield  {journal} {\bibinfo
  {journal} {Rev. Mod. Phys.}\ }\textbf {\bibinfo {volume} {81}},\ \bibinfo
  {pages} {1665} (\bibinfo {year} {2009})}\BibitemShut {NoStop}%
\bibitem [{\citenamefont {Campisi}\ \emph {et~al.}(2011)\citenamefont
  {Campisi}, \citenamefont {H\"anggi},\ and\ \citenamefont
  {Talkner}}]{RevModPhys.83.771}%
  \BibitemOpen
  \bibfield  {author} {\bibinfo {author} {\bibfnamefont {M.}~\bibnamefont
  {Campisi}}, \bibinfo {author} {\bibfnamefont {P.}~\bibnamefont {H\"anggi}},\
  and\ \bibinfo {author} {\bibfnamefont {P.}~\bibnamefont {Talkner}},\
  }\bibfield  {title} {\bibinfo {title} {Colloquium: Quantum fluctuation
  relations: Foundations and applications},\ }\href
  {https://doi.org/10.1103/RevModPhys.83.771} {\bibfield  {journal} {\bibinfo
  {journal} {Rev. Mod. Phys.}\ }\textbf {\bibinfo {volume} {83}},\ \bibinfo
  {pages} {771} (\bibinfo {year} {2011})}\BibitemShut {NoStop}%
\bibitem [{\citenamefont {Goold}\ \emph {et~al.}(2016)\citenamefont {Goold},
  \citenamefont {Huber}, \citenamefont {Riera}, \citenamefont {del Rio},\ and\
  \citenamefont {Skrzypczyk}}]{Goold_2016}%
  \BibitemOpen
  \bibfield  {author} {\bibinfo {author} {\bibfnamefont {J.}~\bibnamefont
  {Goold}}, \bibinfo {author} {\bibfnamefont {M.}~\bibnamefont {Huber}},
  \bibinfo {author} {\bibfnamefont {A.}~\bibnamefont {Riera}}, \bibinfo
  {author} {\bibfnamefont {L.}~\bibnamefont {del Rio}},\ and\ \bibinfo {author}
  {\bibfnamefont {P.}~\bibnamefont {Skrzypczyk}},\ }\bibfield  {title}
  {\bibinfo {title} {The role of quantum information in thermodynamics—a
  topical review},\ }\href {https://doi.org/10.1088/1751-8113/49/14/143001}
  {\bibfield  {journal} {\bibinfo  {journal} {J. Phys. A: Math. Theor.}\
  }\textbf {\bibinfo {volume} {49}},\ \bibinfo {pages} {143001} (\bibinfo
  {year} {2016})}\BibitemShut {NoStop}%
\bibitem [{\citenamefont {Kieu}(2004)}]{PhysRevLett.93.140403}%
  \BibitemOpen
  \bibfield  {author} {\bibinfo {author} {\bibfnamefont {T.~D.}\ \bibnamefont
  {Kieu}},\ }\bibfield  {title} {\bibinfo {title} {The second law, maxwell's
  demon, and work derivable from quantum heat engines},\ }\href
  {https://doi.org/10.1103/PhysRevLett.93.140403} {\bibfield  {journal}
  {\bibinfo  {journal} {Phys. Rev. Lett.}\ }\textbf {\bibinfo {volume} {93}},\
  \bibinfo {pages} {140403} (\bibinfo {year} {2004})}\BibitemShut {NoStop}%
\bibitem [{\citenamefont {Brandão}\ \emph {et~al.}(2015)\citenamefont
  {Brandão}, \citenamefont {Horodecki}, \citenamefont {Ng}, \citenamefont
  {Oppenheim},\ and\ \citenamefont {Wehner}}]{pnas.1411728112}%
  \BibitemOpen
  \bibfield  {author} {\bibinfo {author} {\bibfnamefont {F.}~\bibnamefont
  {Brandão}}, \bibinfo {author} {\bibfnamefont {M.}~\bibnamefont {Horodecki}},
  \bibinfo {author} {\bibfnamefont {N.}~\bibnamefont {Ng}}, \bibinfo {author}
  {\bibfnamefont {J.}~\bibnamefont {Oppenheim}},\ and\ \bibinfo {author}
  {\bibfnamefont {S.}~\bibnamefont {Wehner}},\ }\bibfield  {title} {\bibinfo
  {title} {The second laws of quantum thermodynamics},\ }\href
  {https://doi.org/10.1073/pnas.1411728112} {\bibfield  {journal} {\bibinfo
  {journal} {Proceedings of the National Academy of Sciences}\ }\textbf
  {\bibinfo {volume} {112}},\ \bibinfo {pages} {3275} (\bibinfo {year}
  {2015})}\BibitemShut {NoStop}%
\bibitem [{\citenamefont {Masanes}\ and\ \citenamefont
  {Oppenheim}(2017)}]{WOS:000396104200001}%
  \BibitemOpen
  \bibfield  {author} {\bibinfo {author} {\bibfnamefont {L.}~\bibnamefont
  {Masanes}}\ and\ \bibinfo {author} {\bibfnamefont {J.}~\bibnamefont
  {Oppenheim}},\ }\bibfield  {title} {\bibinfo {title} {A general derivation
  and quantification of the third law of thermodynamics},\ }\href
  {https://doi.org/10.1038/ncomms14538} {\bibfield  {journal} {\bibinfo
  {journal} {Nat. Commun.}\ }\textbf {\bibinfo {volume} {8}},\ \bibinfo {pages}
  {14538} (\bibinfo {year} {2017})}\BibitemShut {NoStop}%
\bibitem [{\citenamefont {Bernardo}(2020)}]{PhysRevE.102.062152}%
  \BibitemOpen
  \bibfield  {author} {\bibinfo {author} {\bibfnamefont {B.~L.}\ \bibnamefont
  {Bernardo}},\ }\bibfield  {title} {\bibinfo {title} {Unraveling the role of
  coherence in the first law of quantum thermodynamics},\ }\href
  {https://doi.org/10.1103/PhysRevE.102.062152} {\bibfield  {journal} {\bibinfo
   {journal} {Phys. Rev. E}\ }\textbf {\bibinfo {volume} {102}},\ \bibinfo
  {pages} {062152} (\bibinfo {year} {2020})}\BibitemShut {NoStop}%
\bibitem [{\citenamefont {Kosloff}(2013)}]{e15062100}%
  \BibitemOpen
  \bibfield  {author} {\bibinfo {author} {\bibfnamefont {R.}~\bibnamefont
  {Kosloff}},\ }\bibfield  {title} {\bibinfo {title} {Quantum thermodynamics: A
  dynamical viewpoint},\ }\href {https://doi.org/10.3390/e15062100} {\bibfield
  {journal} {\bibinfo  {journal} {Entropy}\ }\textbf {\bibinfo {volume} {15}},\
  \bibinfo {pages} {2100} (\bibinfo {year} {2013})}\BibitemShut {NoStop}%
\bibitem [{\citenamefont {\AA{}berg}(2014)}]{PhysRevLett.113.150402}%
  \BibitemOpen
  \bibfield  {author} {\bibinfo {author} {\bibfnamefont {J.}~\bibnamefont
  {\AA{}berg}},\ }\bibfield  {title} {\bibinfo {title} {Catalytic coherence},\
  }\href {https://doi.org/10.1103/PhysRevLett.113.150402} {\bibfield  {journal}
  {\bibinfo  {journal} {Phys. Rev. Lett.}\ }\textbf {\bibinfo {volume} {113}},\
  \bibinfo {pages} {150402} (\bibinfo {year} {2014})}\BibitemShut {NoStop}%
\bibitem [{\citenamefont {Francica}\ \emph {et~al.}(2019)\citenamefont
  {Francica}, \citenamefont {Goold},\ and\ \citenamefont
  {Plastina}}]{PhysRevE.99.042105}%
  \BibitemOpen
  \bibfield  {author} {\bibinfo {author} {\bibfnamefont {G.}~\bibnamefont
  {Francica}}, \bibinfo {author} {\bibfnamefont {J.}~\bibnamefont {Goold}},\
  and\ \bibinfo {author} {\bibfnamefont {F.}~\bibnamefont {Plastina}},\
  }\bibfield  {title} {\bibinfo {title} {Role of coherence in the
  nonequilibrium thermodynamics of quantum systems},\ }\href
  {https://doi.org/10.1103/PhysRevE.99.042105} {\bibfield  {journal} {\bibinfo
  {journal} {Phys. Rev. E}\ }\textbf {\bibinfo {volume} {99}},\ \bibinfo
  {pages} {042105} (\bibinfo {year} {2019})}\BibitemShut {NoStop}%
\bibitem [{\citenamefont {Latune}\ \emph {et~al.}(2019)\citenamefont {Latune},
  \citenamefont {Sinayskiy},\ and\ \citenamefont
  {Petruccione}}]{WOS:000459897600125}%
  \BibitemOpen
  \bibfield  {author} {\bibinfo {author} {\bibfnamefont {C.~L.}\ \bibnamefont
  {Latune}}, \bibinfo {author} {\bibfnamefont {I.}~\bibnamefont {Sinayskiy}},\
  and\ \bibinfo {author} {\bibfnamefont {F.}~\bibnamefont {Petruccione}},\
  }\bibfield  {title} {\bibinfo {title} {Quantum coherence, many-body
  correlations, and non-thermal effects for autonomous thermal machines},\
  }\href {https://doi.org/10.1038/s41598-019-39300-4} {\bibfield  {journal}
  {\bibinfo  {journal} {Sci. Rep.}\ }\textbf {\bibinfo {volume} {9}},\ \bibinfo
  {pages} {3191} (\bibinfo {year} {2019})}\BibitemShut {NoStop}%
\bibitem [{\citenamefont {Lostaglio}\ \emph {et~al.}(2015)\citenamefont
  {Lostaglio}, \citenamefont {Korzekwa}, \citenamefont {Jennings},\ and\
  \citenamefont {Rudolph}}]{PhysRevX.5.021001}%
  \BibitemOpen
  \bibfield  {author} {\bibinfo {author} {\bibfnamefont {M.}~\bibnamefont
  {Lostaglio}}, \bibinfo {author} {\bibfnamefont {K.}~\bibnamefont {Korzekwa}},
  \bibinfo {author} {\bibfnamefont {D.}~\bibnamefont {Jennings}},\ and\
  \bibinfo {author} {\bibfnamefont {T.}~\bibnamefont {Rudolph}},\ }\bibfield
  {title} {\bibinfo {title} {Quantum coherence, time-translation symmetry, and
  thermodynamics},\ }\href {https://doi.org/10.1103/PhysRevX.5.021001}
  {\bibfield  {journal} {\bibinfo  {journal} {Phys. Rev. X}\ }\textbf {\bibinfo
  {volume} {5}},\ \bibinfo {pages} {021001} (\bibinfo {year}
  {2015})}\BibitemShut {NoStop}%
\bibitem [{\citenamefont {Streltsov}\ \emph {et~al.}(2017)\citenamefont
  {Streltsov}, \citenamefont {Adesso},\ and\ \citenamefont
  {Plenio}}]{RevModPhys.89.041003}%
  \BibitemOpen
  \bibfield  {author} {\bibinfo {author} {\bibfnamefont {A.}~\bibnamefont
  {Streltsov}}, \bibinfo {author} {\bibfnamefont {G.}~\bibnamefont {Adesso}},\
  and\ \bibinfo {author} {\bibfnamefont {M.~B.}\ \bibnamefont {Plenio}},\
  }\bibfield  {title} {\bibinfo {title} {Colloquium: Quantum coherence as a
  resource},\ }\href {https://doi.org/10.1103/RevModPhys.89.041003} {\bibfield
  {journal} {\bibinfo  {journal} {Rev. Mod. Phys.}\ }\textbf {\bibinfo {volume}
  {89}},\ \bibinfo {pages} {041003} (\bibinfo {year} {2017})}\BibitemShut
  {NoStop}%
\bibitem [{\citenamefont {Hilt}\ and\ \citenamefont
  {Lutz}(2009)}]{PhysRevA.79.010101}%
  \BibitemOpen
  \bibfield  {author} {\bibinfo {author} {\bibfnamefont {S.}~\bibnamefont
  {Hilt}}\ and\ \bibinfo {author} {\bibfnamefont {E.}~\bibnamefont {Lutz}},\
  }\bibfield  {title} {\bibinfo {title} {System-bath entanglement in quantum
  thermodynamics},\ }\href {https://doi.org/10.1103/PhysRevA.79.010101}
  {\bibfield  {journal} {\bibinfo  {journal} {Phys. Rev. A}\ }\textbf {\bibinfo
  {volume} {79}},\ \bibinfo {pages} {010101} (\bibinfo {year}
  {2009})}\BibitemShut {NoStop}%
\bibitem [{\citenamefont {Santos}\ \emph {et~al.}(2019)\citenamefont {Santos},
  \citenamefont {C{\'e}leri}, \citenamefont {Landi},\ and\ \citenamefont
  {Paternostro}}]{WOS:000462340700001}%
  \BibitemOpen
  \bibfield  {author} {\bibinfo {author} {\bibfnamefont {J.~P.}\ \bibnamefont
  {Santos}}, \bibinfo {author} {\bibfnamefont {L.~C.}\ \bibnamefont
  {C{\'e}leri}}, \bibinfo {author} {\bibfnamefont {G.~T.}\ \bibnamefont
  {Landi}},\ and\ \bibinfo {author} {\bibfnamefont {M.}~\bibnamefont
  {Paternostro}},\ }\bibfield  {title} {\bibinfo {title} {The role of quantum
  coherence in non-equilibrium entropy production},\ }\href
  {https://doi.org/10.1038/s41534-019-0138-y} {\bibfield  {journal} {\bibinfo
  {journal} {npj Quantum Inf.}\ }\textbf {\bibinfo {volume} {5}},\ \bibinfo
  {pages} {23} (\bibinfo {year} {2019})}\BibitemShut {NoStop}%
\bibitem [{\citenamefont {Beny}\ \emph {et~al.}(2018)\citenamefont {Beny},
  \citenamefont {Chubb}, \citenamefont {Farrelly},\ and\ \citenamefont
  {Osborne}}]{WOS:000444757900010}%
  \BibitemOpen
  \bibfield  {author} {\bibinfo {author} {\bibfnamefont {C.}~\bibnamefont
  {Beny}}, \bibinfo {author} {\bibfnamefont {C.~T.}\ \bibnamefont {Chubb}},
  \bibinfo {author} {\bibfnamefont {T.}~\bibnamefont {Farrelly}},\ and\
  \bibinfo {author} {\bibfnamefont {T.~J.}\ \bibnamefont {Osborne}},\
  }\bibfield  {title} {\bibinfo {title} {Energy cost of entanglement extraction
  in complex quantum systems},\ }\href
  {https://doi.org/10.1038/s41467-018-06153-w} {\bibfield  {journal} {\bibinfo
  {journal} {Nat. Commun.}\ }\textbf {\bibinfo {volume} {9}},\ \bibinfo {pages}
  {3792} (\bibinfo {year} {2018})}\BibitemShut {NoStop}%
\bibitem [{\citenamefont {Brandner}\ \emph {et~al.}(2015)\citenamefont
  {Brandner}, \citenamefont {Bauer}, \citenamefont {Schmid},\ and\
  \citenamefont {Seifert}}]{Brandner_2015}%
  \BibitemOpen
  \bibfield  {author} {\bibinfo {author} {\bibfnamefont {K.}~\bibnamefont
  {Brandner}}, \bibinfo {author} {\bibfnamefont {M.}~\bibnamefont {Bauer}},
  \bibinfo {author} {\bibfnamefont {M.~T.}\ \bibnamefont {Schmid}},\ and\
  \bibinfo {author} {\bibfnamefont {U.}~\bibnamefont {Seifert}},\ }\bibfield
  {title} {\bibinfo {title} {Coherence-enhanced efficiency of feedback-driven
  quantum engines},\ }\href {https://doi.org/10.1088/1367-2630/17/6/065006}
  {\bibfield  {journal} {\bibinfo  {journal} {New J. Phys.}\ }\textbf {\bibinfo
  {volume} {17}},\ \bibinfo {pages} {065006} (\bibinfo {year}
  {2015})}\BibitemShut {NoStop}%
\bibitem [{\citenamefont {Korzekwa}\ \emph {et~al.}(2016)\citenamefont
  {Korzekwa}, \citenamefont {Lostaglio}, \citenamefont {Oppenheim},\ and\
  \citenamefont {Jennings}}]{Korzekwa_2016}%
  \BibitemOpen
  \bibfield  {author} {\bibinfo {author} {\bibfnamefont {K.}~\bibnamefont
  {Korzekwa}}, \bibinfo {author} {\bibfnamefont {M.}~\bibnamefont {Lostaglio}},
  \bibinfo {author} {\bibfnamefont {J.}~\bibnamefont {Oppenheim}},\ and\
  \bibinfo {author} {\bibfnamefont {D.}~\bibnamefont {Jennings}},\ }\bibfield
  {title} {\bibinfo {title} {The extraction of work from quantum coherence},\
  }\href {https://doi.org/10.1088/1367-2630/18/2/023045} {\bibfield  {journal}
  {\bibinfo  {journal} {New J. Phys.}\ }\textbf {\bibinfo {volume} {18}},\
  \bibinfo {pages} {023045} (\bibinfo {year} {2016})}\BibitemShut {NoStop}%
\bibitem [{\citenamefont {Breuer}\ and\ \citenamefont
  {Petruccione}(2007)}]{10.1093/acprof:oso/9780199213900.001.0001}%
  \BibitemOpen
  \bibfield  {author} {\bibinfo {author} {\bibfnamefont {H.-P.}\ \bibnamefont
  {Breuer}}\ and\ \bibinfo {author} {\bibfnamefont {F.}~\bibnamefont
  {Petruccione}},\ }\href
  {https://doi.org/10.1093/acprof:oso/9780199213900.001.0001} {\emph {\bibinfo
  {title} {{The Theory of Open Quantum Systems}}}}\ (\bibinfo  {publisher}
  {Oxford University Press},\ \bibinfo {year} {2007})\BibitemShut {NoStop}%
\bibitem [{\citenamefont {Landi}\ and\ \citenamefont
  {Paternostro}(2021)}]{RevModPhys.93.035008}%
  \BibitemOpen
  \bibfield  {author} {\bibinfo {author} {\bibfnamefont {G.~T.}\ \bibnamefont
  {Landi}}\ and\ \bibinfo {author} {\bibfnamefont {M.}~\bibnamefont
  {Paternostro}},\ }\bibfield  {title} {\bibinfo {title} {Irreversible entropy
  production: From classical to quantum},\ }\href
  {https://doi.org/10.1103/RevModPhys.93.035008} {\bibfield  {journal}
  {\bibinfo  {journal} {Rev. Mod. Phys.}\ }\textbf {\bibinfo {volume} {93}},\
  \bibinfo {pages} {035008} (\bibinfo {year} {2021})}\BibitemShut {NoStop}%
\bibitem [{\citenamefont {Weimer}\ \emph {et~al.}(2008)\citenamefont {Weimer},
  \citenamefont {Henrich}, \citenamefont {Rempp}, \citenamefont {Schröder},\
  and\ \citenamefont {Mahler}}]{Weimer_2008}%
  \BibitemOpen
  \bibfield  {author} {\bibinfo {author} {\bibfnamefont {H.}~\bibnamefont
  {Weimer}}, \bibinfo {author} {\bibfnamefont {M.~J.}\ \bibnamefont {Henrich}},
  \bibinfo {author} {\bibfnamefont {F.}~\bibnamefont {Rempp}}, \bibinfo
  {author} {\bibfnamefont {H.}~\bibnamefont {Schröder}},\ and\ \bibinfo
  {author} {\bibfnamefont {G.}~\bibnamefont {Mahler}},\ }\bibfield  {title}
  {\bibinfo {title} {Local effective dynamics of quantum systems: A generalized
  approach to work and heat},\ }\href
  {https://doi.org/10.1209/0295-5075/83/30008} {\bibfield  {journal} {\bibinfo
  {journal} {Europhys. Lett.}\ }\textbf {\bibinfo {volume} {83}},\ \bibinfo
  {pages} {30008} (\bibinfo {year} {2008})}\BibitemShut {NoStop}%
\bibitem [{\citenamefont {Esposito}\ \emph {et~al.}(2010)\citenamefont
  {Esposito}, \citenamefont {Lindenberg},\ and\ \citenamefont {den
  Broeck}}]{Esposito_2010}%
  \BibitemOpen
  \bibfield  {author} {\bibinfo {author} {\bibfnamefont {M.}~\bibnamefont
  {Esposito}}, \bibinfo {author} {\bibfnamefont {K.}~\bibnamefont
  {Lindenberg}},\ and\ \bibinfo {author} {\bibfnamefont {C.~V.}\ \bibnamefont
  {den Broeck}},\ }\bibfield  {title} {\bibinfo {title} {Entropy production as
  correlation between system and reservoir},\ }\href
  {https://doi.org/10.1088/1367-2630/12/1/013013} {\bibfield  {journal}
  {\bibinfo  {journal} {New J. Phys.}\ }\textbf {\bibinfo {volume} {12}},\
  \bibinfo {pages} {013013} (\bibinfo {year} {2010})}\BibitemShut {NoStop}%
\bibitem [{\citenamefont {Teifel}\ and\ \citenamefont
  {Mahler}(2011)}]{PhysRevE.83.041131}%
  \BibitemOpen
  \bibfield  {author} {\bibinfo {author} {\bibfnamefont {J.}~\bibnamefont
  {Teifel}}\ and\ \bibinfo {author} {\bibfnamefont {G.}~\bibnamefont
  {Mahler}},\ }\bibfield  {title} {\bibinfo {title} {Autonomous modular quantum
  systems: Contextual jarzynski relations},\ }\href
  {https://doi.org/10.1103/PhysRevE.83.041131} {\bibfield  {journal} {\bibinfo
  {journal} {Phys. Rev. E}\ }\textbf {\bibinfo {volume} {83}},\ \bibinfo
  {pages} {041131} (\bibinfo {year} {2011})}\BibitemShut {NoStop}%
\bibitem [{\citenamefont {Alipour}\ \emph {et~al.}(2016)\citenamefont
  {Alipour}, \citenamefont {Benatti}, \citenamefont {Bakhshinezhad},
  \citenamefont {Afsary}, \citenamefont {Marcantoni},\ and\ \citenamefont
  {Rezakhani}}]{Alipour2016}%
  \BibitemOpen
  \bibfield  {author} {\bibinfo {author} {\bibfnamefont {S.}~\bibnamefont
  {Alipour}}, \bibinfo {author} {\bibfnamefont {F.}~\bibnamefont {Benatti}},
  \bibinfo {author} {\bibfnamefont {F.}~\bibnamefont {Bakhshinezhad}}, \bibinfo
  {author} {\bibfnamefont {M.}~\bibnamefont {Afsary}}, \bibinfo {author}
  {\bibfnamefont {S.}~\bibnamefont {Marcantoni}},\ and\ \bibinfo {author}
  {\bibfnamefont {A.~T.}\ \bibnamefont {Rezakhani}},\ }\bibfield  {title}
  {\bibinfo {title} {Correlations in quantum thermodynamics: Heat, work, and
  entropy production},\ }\href {https://doi.org/10.1038/srep35568} {\bibfield
  {journal} {\bibinfo  {journal} {Sci. Rep.}\ }\textbf {\bibinfo {volume}
  {6}},\ \bibinfo {pages} {35568} (\bibinfo {year} {2016})}\BibitemShut
  {NoStop}%
\bibitem [{\citenamefont {Strasberg}\ \emph {et~al.}(2017)\citenamefont
  {Strasberg}, \citenamefont {Schaller}, \citenamefont {Brandes},\ and\
  \citenamefont {Esposito}}]{PhysRevX.7.021003}%
  \BibitemOpen
  \bibfield  {author} {\bibinfo {author} {\bibfnamefont {P.}~\bibnamefont
  {Strasberg}}, \bibinfo {author} {\bibfnamefont {G.}~\bibnamefont {Schaller}},
  \bibinfo {author} {\bibfnamefont {T.}~\bibnamefont {Brandes}},\ and\ \bibinfo
  {author} {\bibfnamefont {M.}~\bibnamefont {Esposito}},\ }\bibfield  {title}
  {\bibinfo {title} {Quantum and information thermodynamics: A unifying
  framework based on repeated interactions},\ }\href
  {https://doi.org/10.1103/PhysRevX.7.021003} {\bibfield  {journal} {\bibinfo
  {journal} {Phys. Rev. X}\ }\textbf {\bibinfo {volume} {7}},\ \bibinfo {pages}
  {021003} (\bibinfo {year} {2017})}\BibitemShut {NoStop}%
\bibitem [{\citenamefont {Popovic}\ \emph {et~al.}(2018)\citenamefont
  {Popovic}, \citenamefont {Vacchini},\ and\ \citenamefont
  {Campbell}}]{PhysRevA.98.012130}%
  \BibitemOpen
  \bibfield  {author} {\bibinfo {author} {\bibfnamefont {M.}~\bibnamefont
  {Popovic}}, \bibinfo {author} {\bibfnamefont {B.}~\bibnamefont {Vacchini}},\
  and\ \bibinfo {author} {\bibfnamefont {S.}~\bibnamefont {Campbell}},\
  }\bibfield  {title} {\bibinfo {title} {Entropy production and correlations in
  a controlled non-markovian setting},\ }\href
  {https://doi.org/10.1103/PhysRevA.98.012130} {\bibfield  {journal} {\bibinfo
  {journal} {Phys. Rev. A}\ }\textbf {\bibinfo {volume} {98}},\ \bibinfo
  {pages} {012130} (\bibinfo {year} {2018})}\BibitemShut {NoStop}%
\bibitem [{\citenamefont {Strasberg}\ and\ \citenamefont
  {Esposito}(2019)}]{PhysRevE.99.012120}%
  \BibitemOpen
  \bibfield  {author} {\bibinfo {author} {\bibfnamefont {P.}~\bibnamefont
  {Strasberg}}\ and\ \bibinfo {author} {\bibfnamefont {M.}~\bibnamefont
  {Esposito}},\ }\bibfield  {title} {\bibinfo {title} {Non-markovianity and
  negative entropy production rates},\ }\href
  {https://doi.org/10.1103/PhysRevE.99.012120} {\bibfield  {journal} {\bibinfo
  {journal} {Phys. Rev. E}\ }\textbf {\bibinfo {volume} {99}},\ \bibinfo
  {pages} {012120} (\bibinfo {year} {2019})}\BibitemShut {NoStop}%
\bibitem [{\citenamefont {Rivas}(2020)}]{PhysRevLett.124.160601}%
  \BibitemOpen
  \bibfield  {author} {\bibinfo {author} {\bibfnamefont {A.}~\bibnamefont
  {Rivas}},\ }\bibfield  {title} {\bibinfo {title} {Strong coupling
  thermodynamics of open quantum systems},\ }\href
  {https://doi.org/10.1103/PhysRevLett.124.160601} {\bibfield  {journal}
  {\bibinfo  {journal} {Phys. Rev. Lett.}\ }\textbf {\bibinfo {volume} {124}},\
  \bibinfo {pages} {160601} (\bibinfo {year} {2020})}\BibitemShut {NoStop}%
\bibitem [{\citenamefont {Alipour}\ \emph {et~al.}(2022)\citenamefont
  {Alipour}, \citenamefont {Rezakhani}, \citenamefont {Chenu}, \citenamefont
  {del Campo},\ and\ \citenamefont {Ala-Nissila}}]{PhysRevA.105.L040201}%
  \BibitemOpen
  \bibfield  {author} {\bibinfo {author} {\bibfnamefont {S.}~\bibnamefont
  {Alipour}}, \bibinfo {author} {\bibfnamefont {A.~T.}\ \bibnamefont
  {Rezakhani}}, \bibinfo {author} {\bibfnamefont {A.}~\bibnamefont {Chenu}},
  \bibinfo {author} {\bibfnamefont {A.}~\bibnamefont {del Campo}},\ and\
  \bibinfo {author} {\bibfnamefont {T.}~\bibnamefont {Ala-Nissila}},\
  }\bibfield  {title} {\bibinfo {title} {Entropy-based formulation of
  thermodynamics in arbitrary quantum evolution},\ }\href
  {https://doi.org/10.1103/PhysRevA.105.L040201} {\bibfield  {journal}
  {\bibinfo  {journal} {Phys. Rev. A}\ }\textbf {\bibinfo {volume} {105}},\
  \bibinfo {pages} {L040201} (\bibinfo {year} {2022})}\BibitemShut {NoStop}%
\bibitem [{\citenamefont {Colla}\ and\ \citenamefont
  {Breuer}(2022)}]{PhysRevA.105.052216}%
  \BibitemOpen
  \bibfield  {author} {\bibinfo {author} {\bibfnamefont {A.}~\bibnamefont
  {Colla}}\ and\ \bibinfo {author} {\bibfnamefont {H.-P.}\ \bibnamefont
  {Breuer}},\ }\bibfield  {title} {\bibinfo {title} {Open-system approach to
  nonequilibrium quantum thermodynamics at arbitrary coupling},\ }\href
  {https://doi.org/10.1103/PhysRevA.105.052216} {\bibfield  {journal} {\bibinfo
   {journal} {Phys. Rev. A}\ }\textbf {\bibinfo {volume} {105}},\ \bibinfo
  {pages} {052216} (\bibinfo {year} {2022})}\BibitemShut {NoStop}%
\bibitem [{\citenamefont {Bender}\ \emph {et~al.}(2000)\citenamefont {Bender},
  \citenamefont {Brody},\ and\ \citenamefont {Meister}}]{Bender_2000}%
  \BibitemOpen
  \bibfield  {author} {\bibinfo {author} {\bibfnamefont {C.~M.}\ \bibnamefont
  {Bender}}, \bibinfo {author} {\bibfnamefont {D.~C.}\ \bibnamefont {Brody}},\
  and\ \bibinfo {author} {\bibfnamefont {B.~K.}\ \bibnamefont {Meister}},\
  }\bibfield  {title} {\bibinfo {title} {Quantum mechanical carnot engine},\
  }\href {https://doi.org/10.1088/0305-4470/33/24/302} {\bibfield  {journal}
  {\bibinfo  {journal} {J. Phys. A: Math. Gen.}\ }\textbf {\bibinfo {volume}
  {33}},\ \bibinfo {pages} {4427} (\bibinfo {year} {2000})}\BibitemShut
  {NoStop}%
\bibitem [{\citenamefont {Abe}(2011)}]{PhysRevE.83.041117}%
  \BibitemOpen
  \bibfield  {author} {\bibinfo {author} {\bibfnamefont {S.}~\bibnamefont
  {Abe}},\ }\bibfield  {title} {\bibinfo {title} {Maximum-power
  quantum-mechanical carnot engine},\ }\href
  {https://doi.org/10.1103/PhysRevE.83.041117} {\bibfield  {journal} {\bibinfo
  {journal} {Phys. Rev. E}\ }\textbf {\bibinfo {volume} {83}},\ \bibinfo
  {pages} {041117} (\bibinfo {year} {2011})}\BibitemShut {NoStop}%
\bibitem [{\citenamefont {Polkovnikov}(2011)}]{POLKOVNIKOV2011486}%
  \BibitemOpen
  \bibfield  {author} {\bibinfo {author} {\bibfnamefont {A.}~\bibnamefont
  {Polkovnikov}},\ }\bibfield  {title} {\bibinfo {title} {Microscopic diagonal
  entropy and its connection to basic thermodynamic relations},\ }\href
  {https://doi.org/https://doi.org/10.1016/j.aop.2010.08.004} {\bibfield
  {journal} {\bibinfo  {journal} {Ann. Phys. (NY)}\ }\textbf {\bibinfo {volume}
  {326}},\ \bibinfo {pages} {486} (\bibinfo {year} {2011})}\BibitemShut
  {NoStop}%
\bibitem [{\citenamefont {Kieu}(2006)}]{Kieu2006}%
  \BibitemOpen
  \bibfield  {author} {\bibinfo {author} {\bibfnamefont {T.~D.}\ \bibnamefont
  {Kieu}},\ }\bibfield  {title} {\bibinfo {title} {Quantum heat engines, the
  second law and maxwell's daemon},\ }\href
  {https://doi.org/10.1140/epjd/e2006-00075-5} {\bibfield  {journal} {\bibinfo
  {journal} {Eur. J. Phys. D}\ }\textbf {\bibinfo {volume} {39}},\ \bibinfo
  {pages} {115} (\bibinfo {year} {2006})}\BibitemShut {NoStop}%
\bibitem [{\citenamefont {Quan}\ \emph {et~al.}(2007)\citenamefont {Quan},
  \citenamefont {Liu}, \citenamefont {Sun},\ and\ \citenamefont
  {Nori}}]{PhysRevE.76.031105}%
  \BibitemOpen
  \bibfield  {author} {\bibinfo {author} {\bibfnamefont {H.~T.}\ \bibnamefont
  {Quan}}, \bibinfo {author} {\bibfnamefont {Y.-X.}\ \bibnamefont {Liu}},
  \bibinfo {author} {\bibfnamefont {C.~P.}\ \bibnamefont {Sun}},\ and\ \bibinfo
  {author} {\bibfnamefont {F.}~\bibnamefont {Nori}},\ }\bibfield  {title}
  {\bibinfo {title} {Quantum thermodynamic cycles and quantum heat engines},\
  }\href {https://doi.org/10.1103/PhysRevE.76.031105} {\bibfield  {journal}
  {\bibinfo  {journal} {Phys. Rev. E}\ }\textbf {\bibinfo {volume} {76}},\
  \bibinfo {pages} {031105} (\bibinfo {year} {2007})}\BibitemShut {NoStop}%
\bibitem [{\citenamefont {Alicki}(1979)}]{Alicki_1979}%
  \BibitemOpen
  \bibfield  {author} {\bibinfo {author} {\bibfnamefont {R.}~\bibnamefont
  {Alicki}},\ }\bibfield  {title} {\bibinfo {title} {The quantum open system as
  a model of the heat engine},\ }\href
  {https://doi.org/10.1088/0305-4470/12/5/007} {\bibfield  {journal} {\bibinfo
  {journal} {J. Phys. A: Math. Theor.}\ }\textbf {\bibinfo {volume} {12}},\
  \bibinfo {pages} {L103} (\bibinfo {year} {1979})}\BibitemShut {NoStop}%
\bibitem [{\citenamefont {Spohn}(2008)}]{10.1063/1.523789}%
  \BibitemOpen
  \bibfield  {author} {\bibinfo {author} {\bibfnamefont {H.}~\bibnamefont
  {Spohn}},\ }\bibfield  {title} {\bibinfo {title} {Entropy production for
  quantum dynamical semigroups},\ }\href {https://doi.org/10.1063/1.523789}
  {\bibfield  {journal} {\bibinfo  {journal} {J. Math. Phys.}\ }\textbf
  {\bibinfo {volume} {19}},\ \bibinfo {pages} {1227} (\bibinfo {year}
  {2008})}\BibitemShut {NoStop}%
\bibitem [{\citenamefont {Talkner}\ \emph {et~al.}(2007)\citenamefont
  {Talkner}, \citenamefont {Lutz},\ and\ \citenamefont
  {H\"anggi}}]{PhysRevE.75.050102}%
  \BibitemOpen
  \bibfield  {author} {\bibinfo {author} {\bibfnamefont {P.}~\bibnamefont
  {Talkner}}, \bibinfo {author} {\bibfnamefont {E.}~\bibnamefont {Lutz}},\ and\
  \bibinfo {author} {\bibfnamefont {P.}~\bibnamefont {H\"anggi}},\ }\bibfield
  {title} {\bibinfo {title} {Fluctuation theorems: Work is not an observable},\
  }\href {https://doi.org/10.1103/PhysRevE.75.050102} {\bibfield  {journal}
  {\bibinfo  {journal} {Phys. Rev. E}\ }\textbf {\bibinfo {volume} {75}},\
  \bibinfo {pages} {050102} (\bibinfo {year} {2007})}\BibitemShut {NoStop}%
\bibitem [{\citenamefont {Vilar}\ and\ \citenamefont
  {Rubi}(2008)}]{PhysRevLett.100.020601}%
  \BibitemOpen
  \bibfield  {author} {\bibinfo {author} {\bibfnamefont {J.~M.~G.}\
  \bibnamefont {Vilar}}\ and\ \bibinfo {author} {\bibfnamefont {J.~M.}\
  \bibnamefont {Rubi}},\ }\bibfield  {title} {\bibinfo {title} {Failure of the
  work-hamiltonian connection for free-energy calculations},\ }\href
  {https://doi.org/10.1103/PhysRevLett.100.020601} {\bibfield  {journal}
  {\bibinfo  {journal} {Phys. Rev. Lett.}\ }\textbf {\bibinfo {volume} {100}},\
  \bibinfo {pages} {020601} (\bibinfo {year} {2008})}\BibitemShut {NoStop}%
\bibitem [{\citenamefont {Gelbwaser-Klimovsky}\ and\ \citenamefont
  {Aspuru-Guzik}(2017)}]{C6SC04350J}%
  \BibitemOpen
  \bibfield  {author} {\bibinfo {author} {\bibfnamefont {D.}~\bibnamefont
  {Gelbwaser-Klimovsky}}\ and\ \bibinfo {author} {\bibfnamefont
  {A.}~\bibnamefont {Aspuru-Guzik}},\ }\bibfield  {title} {\bibinfo {title} {On
  thermodynamic inconsistencies in several photosynthetic and solar cell models
  and how to fix them},\ }\href {https://doi.org/10.1039/C6SC04350J} {\bibfield
   {journal} {\bibinfo  {journal} {Chem. Sci.}\ }\textbf {\bibinfo {volume}
  {8}},\ \bibinfo {pages} {1008} (\bibinfo {year} {2017})}\BibitemShut
  {NoStop}%
\bibitem [{\citenamefont {Niedenzu}\ \emph {et~al.}(2019)\citenamefont
  {Niedenzu}, \citenamefont {Huber},\ and\ \citenamefont
  {Boukobza}}]{Niedenzu2019conceptsofworkin}%
  \BibitemOpen
  \bibfield  {author} {\bibinfo {author} {\bibfnamefont {W.}~\bibnamefont
  {Niedenzu}}, \bibinfo {author} {\bibfnamefont {M.}~\bibnamefont {Huber}},\
  and\ \bibinfo {author} {\bibfnamefont {E.}~\bibnamefont {Boukobza}},\
  }\bibfield  {title} {\bibinfo {title} {Concepts of work in autonomous quantum
  heat engines},\ }\href {https://doi.org/10.22331/q-2019-10-14-195} {\bibfield
   {journal} {\bibinfo  {journal} {{Quantum}}\ }\textbf {\bibinfo {volume}
  {3}},\ \bibinfo {pages} {195} (\bibinfo {year} {2019})}\BibitemShut {NoStop}%
\bibitem [{\citenamefont {Sampaio}\ \emph {et~al.}(2018)\citenamefont
  {Sampaio}, \citenamefont {Suomela}, \citenamefont {Ala-Nissila},
  \citenamefont {Anders},\ and\ \citenamefont {Philbin}}]{PhysRevA.97.012131}%
  \BibitemOpen
  \bibfield  {author} {\bibinfo {author} {\bibfnamefont {R.}~\bibnamefont
  {Sampaio}}, \bibinfo {author} {\bibfnamefont {S.}~\bibnamefont {Suomela}},
  \bibinfo {author} {\bibfnamefont {T.}~\bibnamefont {Ala-Nissila}}, \bibinfo
  {author} {\bibfnamefont {J.}~\bibnamefont {Anders}},\ and\ \bibinfo {author}
  {\bibfnamefont {T.~G.}\ \bibnamefont {Philbin}},\ }\bibfield  {title}
  {\bibinfo {title} {Quantum work in the bohmian framework},\ }\href
  {https://doi.org/10.1103/PhysRevA.97.012131} {\bibfield  {journal} {\bibinfo
  {journal} {Phys. Rev. A}\ }\textbf {\bibinfo {volume} {97}},\ \bibinfo
  {pages} {012131} (\bibinfo {year} {2018})}\BibitemShut {NoStop}%
\bibitem [{\citenamefont {Elouard}\ \emph {et~al.}(2017)\citenamefont
  {Elouard}, \citenamefont {Herrera-Mart{\'\i}}, \citenamefont {Clusel},\ and\
  \citenamefont {Auff{\`e}ves}}]{Elouard2017}%
  \BibitemOpen
  \bibfield  {author} {\bibinfo {author} {\bibfnamefont {C.}~\bibnamefont
  {Elouard}}, \bibinfo {author} {\bibfnamefont {D.~A.}\ \bibnamefont
  {Herrera-Mart{\'\i}}}, \bibinfo {author} {\bibfnamefont {M.}~\bibnamefont
  {Clusel}},\ and\ \bibinfo {author} {\bibfnamefont {A.}~\bibnamefont
  {Auff{\`e}ves}},\ }\bibfield  {title} {\bibinfo {title} {The role of quantum
  measurement in stochastic thermodynamics},\ }\href
  {https://doi.org/10.1038/s41534-017-0008-4} {\bibfield  {journal} {\bibinfo
  {journal} {npj Quantum Information}\ }\textbf {\bibinfo {volume} {3}},\
  \bibinfo {pages} {9} (\bibinfo {year} {2017})}\BibitemShut {NoStop}%
\bibitem [{\citenamefont {Neves}\ and\ \citenamefont
  {Brito}(2023)}]{PhysRevA.108.042209}%
  \BibitemOpen
  \bibfield  {author} {\bibinfo {author} {\bibfnamefont {L.~R.~T.}\
  \bibnamefont {Neves}}\ and\ \bibinfo {author} {\bibfnamefont
  {F.}~\bibnamefont {Brito}},\ }\bibfield  {title} {\bibinfo {title}
  {Constraint on local definitions of quantum internal energy},\ }\href
  {https://doi.org/10.1103/PhysRevA.108.042209} {\bibfield  {journal} {\bibinfo
   {journal} {Phys. Rev. A}\ }\textbf {\bibinfo {volume} {108}},\ \bibinfo
  {pages} {042209} (\bibinfo {year} {2023})}\BibitemShut {NoStop}%
\bibitem [{\citenamefont {Schlienz}\ and\ \citenamefont
  {Mahler}(1995)}]{PhysRevA.52.4396}%
  \BibitemOpen
  \bibfield  {author} {\bibinfo {author} {\bibfnamefont {J.}~\bibnamefont
  {Schlienz}}\ and\ \bibinfo {author} {\bibfnamefont {G.}~\bibnamefont
  {Mahler}},\ }\bibfield  {title} {\bibinfo {title} {Description of
  entanglement},\ }\href {https://doi.org/10.1103/PhysRevA.52.4396} {\bibfield
  {journal} {\bibinfo  {journal} {Phys. Rev. A}\ }\textbf {\bibinfo {volume}
  {52}},\ \bibinfo {pages} {4396} (\bibinfo {year} {1995})}\BibitemShut
  {NoStop}%
\bibitem [{\citenamefont {Carrega}\ \emph {et~al.}(2016)\citenamefont
  {Carrega}, \citenamefont {Solinas}, \citenamefont {Sassetti},\ and\
  \citenamefont {Weiss}}]{PhysRevLett.116.240403}%
  \BibitemOpen
  \bibfield  {author} {\bibinfo {author} {\bibfnamefont {M.}~\bibnamefont
  {Carrega}}, \bibinfo {author} {\bibfnamefont {P.}~\bibnamefont {Solinas}},
  \bibinfo {author} {\bibfnamefont {M.}~\bibnamefont {Sassetti}},\ and\
  \bibinfo {author} {\bibfnamefont {U.}~\bibnamefont {Weiss}},\ }\bibfield
  {title} {\bibinfo {title} {Energy exchange in driven open quantum systems at
  strong coupling},\ }\href {https://doi.org/10.1103/PhysRevLett.116.240403}
  {\bibfield  {journal} {\bibinfo  {journal} {Phys. Rev. Lett.}\ }\textbf
  {\bibinfo {volume} {116}},\ \bibinfo {pages} {240403} (\bibinfo {year}
  {2016})}\BibitemShut {NoStop}%
\bibitem [{\citenamefont {Bernardo}(2021)}]{PhysRevE.104.044111}%
  \BibitemOpen
  \bibfield  {author} {\bibinfo {author} {\bibfnamefont {B.~d.~L.}\
  \bibnamefont {Bernardo}},\ }\bibfield  {title} {\bibinfo {title} {Relating
  heat and entanglement in strong-coupling thermodynamics},\ }\href
  {https://doi.org/10.1103/PhysRevE.104.044111} {\bibfield  {journal} {\bibinfo
   {journal} {Phys. Rev. E}\ }\textbf {\bibinfo {volume} {104}},\ \bibinfo
  {pages} {044111} (\bibinfo {year} {2021})}\BibitemShut {NoStop}%
\bibitem [{\citenamefont {Singh}\ \emph {et~al.}(2003)\citenamefont {Singh},
  \citenamefont {Tong}, \citenamefont {Basu}, \citenamefont {Chen},\ and\
  \citenamefont {Du}}]{PhysRevA.67.032106}%
  \BibitemOpen
  \bibfield  {author} {\bibinfo {author} {\bibfnamefont {K.}~\bibnamefont
  {Singh}}, \bibinfo {author} {\bibfnamefont {D.~M.}\ \bibnamefont {Tong}},
  \bibinfo {author} {\bibfnamefont {K.}~\bibnamefont {Basu}}, \bibinfo {author}
  {\bibfnamefont {J.~L.}\ \bibnamefont {Chen}},\ and\ \bibinfo {author}
  {\bibfnamefont {J.~F.}\ \bibnamefont {Du}},\ }\bibfield  {title} {\bibinfo
  {title} {Geometric phases for nondegenerate and degenerate mixed states},\
  }\href {https://doi.org/10.1103/PhysRevA.67.032106} {\bibfield  {journal}
  {\bibinfo  {journal} {Phys. Rev. A}\ }\textbf {\bibinfo {volume} {67}},\
  \bibinfo {pages} {032106} (\bibinfo {year} {2003})}\BibitemShut {NoStop}%
\bibitem [{\citenamefont {Tong}\ \emph {et~al.}(2004)\citenamefont {Tong},
  \citenamefont {Sj\"oqvist}, \citenamefont {Kwek},\ and\ \citenamefont
  {Oh}}]{PhysRevLett.93.080405}%
  \BibitemOpen
  \bibfield  {author} {\bibinfo {author} {\bibfnamefont {D.~M.}\ \bibnamefont
  {Tong}}, \bibinfo {author} {\bibfnamefont {E.}~\bibnamefont {Sj\"oqvist}},
  \bibinfo {author} {\bibfnamefont {L.~C.}\ \bibnamefont {Kwek}},\ and\
  \bibinfo {author} {\bibfnamefont {C.~H.}\ \bibnamefont {Oh}},\ }\bibfield
  {title} {\bibinfo {title} {Kinematic approach to the mixed state geometric
  phase in nonunitary evolution},\ }\href
  {https://doi.org/10.1103/PhysRevLett.93.080405} {\bibfield  {journal}
  {\bibinfo  {journal} {Phys. Rev. Lett.}\ }\textbf {\bibinfo {volume} {93}},\
  \bibinfo {pages} {080405} (\bibinfo {year} {2004})}\BibitemShut {NoStop}%
\bibitem [{\citenamefont {Nicacio}\ and\ \citenamefont
  {Maia}(2023)}]{PhysRevA.108.022209}%
  \BibitemOpen
  \bibfield  {author} {\bibinfo {author} {\bibfnamefont {F.}~\bibnamefont
  {Nicacio}}\ and\ \bibinfo {author} {\bibfnamefont {R.~N.~P.}\ \bibnamefont
  {Maia}},\ }\bibfield  {title} {\bibinfo {title} {Gauge quantum thermodynamics
  of time-local non-markovian evolutions},\ }\href
  {https://doi.org/10.1103/PhysRevA.108.022209} {\bibfield  {journal} {\bibinfo
   {journal} {Phys. Rev. A}\ }\textbf {\bibinfo {volume} {108}},\ \bibinfo
  {pages} {022209} (\bibinfo {year} {2023})}\BibitemShut {NoStop}%
\bibitem [{\citenamefont {Pu}\ and\ \citenamefont {Zhou}()}]{Tobe}%
  \BibitemOpen
  \bibfield  {author} {\bibinfo {author} {\bibfnamefont {J.}~\bibnamefont
  {Pu}}\ and\ \bibinfo {author} {\bibfnamefont {T.}~\bibnamefont {Zhou}},\
  }\href@noop {} {\bibinfo {title} {Unitary and dissipative dynamics of an open
  quantum system, to be submitted.}}\BibitemShut {Stop}%
\end{thebibliography}%

\end{document}